\documentclass[iop,apjl,numberedappendix,twocolappendix,revtex4]{emulateapj}
\usepackage{setspace}
\usepackage{amsfonts,amsmath,amssymb}
\usepackage{txfonts}
\usepackage{graphicx}
\usepackage{bm}
\usepackage{epstopdf}
\usepackage{verbatim}
\usepackage{xcolor}
\usepackage{units}
\usepackage[T1]{fontenc}
\usepackage[latin9]{inputenc}
\usepackage{amsbsy}
\usepackage{esint}

\begin{document}
\title{A stringent limit on the amplitude of Alfv\'enic perturbations in high-beta low-collisionality plasmas}

\author{J.~Squire}
\email{jsquire@caltech.edu}
\affiliation{Theoretical Astrophysics, 350-17, California Institute of Technology, Pasadena, CA 91125, USA}
\affiliation{Walter Burke Institute for Theoretical Physics}
\author{E.~Quataert}
\affiliation{Astronomy Department and Theoretical Astrophysics Center, University of California, Berkeley, CA 94720, USA}
\author{A.~A.~Schekochihin}
\affiliation{The Rudolf Peierls Centre for Theoretical Physics, University of Oxford, 1 Keble Road, Oxford, OX1 3NP, UK}
\affiliation{Merton College, Oxford OX1 4JD, UK}

\begin{abstract}
It is shown that low-collisionality plasmas cannot support linearly polarized shear-Alfv\'en fluctuations above a 
critical amplitude $\delta B_{\perp}/B_{0} \sim \beta^{\,-1/2}$, where $\beta$ is the ratio of thermal  to magnetic pressure. 
Above this cutoff, a developing fluctuation will generate a pressure anisotropy that is 
sufficient to destabilize itself through the parallel firehose instability. This causes the wave 
frequency to approach zero, interrupting the fluctuation before any oscillation. 
The  magnetic field lines rapidly relax into a sequence of  angular zig-zag structures. Such a
 restrictive bound on shear-Alfv\'en-wave amplitudes has far-reaching implications for the physics of magnetized turbulence
 in the high-$\beta$ conditions prevalent in many astrophysical plasmas, as
well as for  the solar wind at $\sim 1 \mathrm{AU}$ where $\beta \gtrsim 1$.
\end{abstract}
\maketitle


\section{Introduction}

Shear-Alfv\'en waves are perhaps the most fundamental of all oscillations in a magnetized plasma \citep{Alfven:1942hl}.
Their existence provides a key distinction 
between neutral and magnetized fluids, and they play a central theoretical role
 in most sub-disciplines of plasma physics, 
 including magnetized turbulence 
 \citep{Goldreich:1995hq,Ng:1996ik}, the solar wind \citep{Ofman:2010jg,Bruno:2013hk}, the solar corona \citep{Marsch:2006vp} and magnetic fusion \citep{Heidbrink:2008eg}.
This general applicability has led to 
 intense study of their basic properties \citep{Cramer:2011uc}.
This research -- which includes studies ranging from kinetic physics and damping mechanisms \citep{Foote:1979ee}, to nonlinear
instabilities \citep{Medvedev:1997dk,Hamabata:1993jk} and the effects of inhomogeneity \citep{Velli:1993to} -- has in turn been 
vital for the formulation of more applied theories.
Interestingly, the low-frequency shear-Alfv\'en wave specifically has emerged relatively 
unscathed from this onslaught of theoretical inquiry (but see \citealt{Delzanna:2001,Cramer:2011uc,Bruno:2013hk} and references therein), 
apparently  being much less affected by kinetic damping mechanisms and 
other nonideal effects  than its fast and slow wave cousins \citep{Foote:1979ee,Schekochihin:2007dm}.

In this Letter, we discuss a dramatic departure from this behavior, showing that a high-beta collisionless plasma 
cannot support linearly polarized shear-Alfv\'en (SA) fluctuations above the critical amplitude,
\begin{equation}
(\delta B_{\perp} /B_{0})_{\mathrm{max}} \sim  \beta^{\,-1/2},\label{eq:lim}
\end{equation}
where $\beta \equiv 8\pi p_{0}/B_{0}^{2}$ is  the ratio of thermal  pressure to
magnetic pressure.
This upper bound is independent of the spatial scale of the perturbation (as long as it is above the ion Larmor radius), and
a similar restriction also holds in the weakly collisional Braginskii  limit \citep{Braginskii:1965vl}.
For fluctuations with $\delta B_{\perp}/B_{0} \gtrsim\beta^{\,-1/2}$, the magnetic field 
rapidly forms a sequence of zig-zags -- straight field line segments joined by sharp corners -- 
maintaining this configuration with the magnetic energy far in excess of the kinetic energy.

What is the cause of such dramatic nonlinear behavior, even in regimes ($\delta B_{\perp}/B_{0}\ll1$ for $\beta\gg 1$) where linear physics might appear to be applicable?
In a magnetized plasma in which the ion gyro-frequency $\Omega_{c}$ is much larger than the collision frequency $\nu_{c}$, a decreasing (in time) magnetic field leads -- due to conservation of particle magnetic moment $\mu = m v_{\perp}^{2}/2B$ --
to a decreasing pressure perpendicular to the magnetic field ($p_{\perp}$), while the parallel pressure ($p_{\parallel}$) increases.
This anisotropy, $\Delta p \equiv p_{\perp}-p_{\parallel}<0$, neutralizes the restoring effects of magnetic tension,  destabilizing the SA wave if $\Delta p  < - {B^{2}}/{4\pi}$.
This well-studied instability is known as the {parallel firehose} \citep{Rosenbluth:1956,Chandrasekhar:1958,Parker:1958,Schekochihin:2010bv}. Now consider the ensuing dynamics if 
we start with $\Delta p =0$, but with a field that, in the process
of decreasing due to the Lorentz force,  \emph{generates a pressure anisotropy that would be sufficient to 
destabilize itself}. This is a nonlinear effect not captured in linear models of SA waves.
As $\Delta p$ approaches the firehose limit, the magnetic tension disappears and the Alfv\'en frequency approaches 
zero, interrupting the development of the wave.  As shown below, because the wave perturbs the field magnitude 
by  $\delta B_{\perp}^{2}$, an amplitude 
 $\delta B_{\perp} /B_{0} \gtrsim \beta^{\,-1/2}$ is sufficient
to generate such a $\Delta p$ in a collisionless plasma. 
As the field decrease is interrupted at the firehose stability boundary, the plasma self-organizes to 
prevent  further changes in field strength, leading to the development of piecewise-straight (and therefore, tension-less) field-line structures.

This Letter explores the physics of this stringent amplitude limit, starting with simple
analytic considerations. We then numerically examine the nonlinear behavior of fluctuations with amplitudes
that exceed the limit  and conclude
with a discussion of possible implications for astrophysical turbulence and the solar wind.
We focus primarily on the fate of an isolated $B_{\perp}$ perturbation -- i.e., a linearly polarized standing wave -- because this 
case is the simplest physically. Both the amplitude limit itself, and the plasma dynamics as the system approaches the firehose limit, are similar for traveling waves and for an initial velocity perturbation. Circularly polarized perturbations are, however, unaffected. 

\section{Theory} 
\begin{figure}
\begin{center}
\includegraphics[width=1.0\columnwidth]{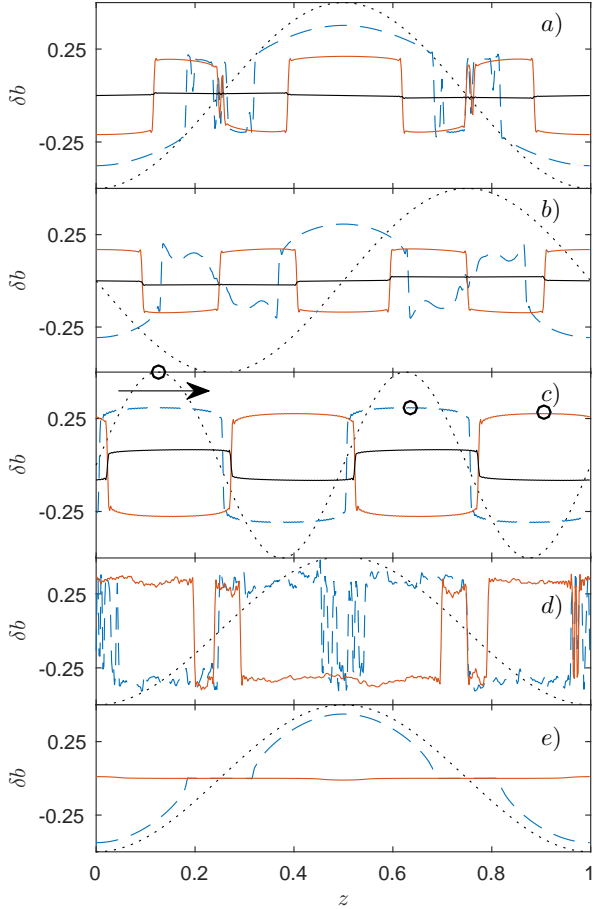}
\caption{Evolution of $\delta b={B}_{x}/B_{z0}$ in a $\beta = 100$ plasma. (a)--(c) show  solutions
of the full collisionless LF equations \eqref{eq:KMHD rho}-\eqref{eq:KMHD pl} in one dimension, starting from (a) $\delta b(0)= -0.5 \cos(2\pi z)$, (b) $
u_{x}(0) =-0.5 v_{A} \sin(2\pi z)$ [linearly, this $u_{x}(0)$ leads to $\delta b= -0.5 \cos(2\pi z)$], and (c) a traveling wave $\delta b(0)=-u_{x}(0)/v_{A}= 0.5 \sin(4\pi z)$. (d) and (e) show standing-wave solutions
of the nonlinear wave equation
\eqref{eq:waves}, with (d) the collisionless closure \eqref{eq:noC Delta}, and (e) the
Braginskii closure \eqref{eq:brag Delta} (with $\omega_{A}\, \beta /\nu_{c} = 100$).
Each solution uses 512 Fourier modes. The figures show $\delta b$ at $t=0$ [black dotted line; $u_{x}(0)$ is shown in (b)], $\delta b$ at $t=\tau_{A}/2$ [blue dashed line; $t=3\tau_{A}/2$ in (c)], $\delta b$ at $t=3 \tau_{A}$ (red solid line), and $u_{x}/v_{A}$ at $t=3 \tau_{A}$ [black solid line; only in (a)--(c)], where $\tau_{A} = 2\pi /\omega_{A}$. The circles in (c) show the same position on the wave as it evolves, illustrating
its decrease in speed as the wave damps. Note the 
strong damping of velocity at late times in (a)--(c) [the wave is not fully interrupted by 
the final time shown in (c)], and the decay 
of the perturbation to $\delta b < (\delta b_{0})_{\mathrm{max}}$ by $t=3 \tau_{A}$ in (e). 
The highly nonlinear behavior in each case shown here starkly contrasts with
the almost perfectly linear evolution of an MHD SA fluctuation at these 
parameters.
}
\label{fig:1}
\end{center}
\end{figure}

On spatiotemporal scales  larger than those 
relating to particle gyromotion, the particle distribution function is approximately gyrotropic.
The magnetic field and first three moments of the kinetic equation then satisfy  \citep{Kulsrud:1980tm,Schekochihin:2010bv}
\begin{align}
&\partial_{t}\rho +\nabla \cdot (\rho \bm{u} ) = 0,\label{eq:KMHD rho} \\[2ex]
\rho \left(\partial_{t}\bm{u} +  \bm{u}\cdot \nabla \bm{u} \right)= -& \nabla\left( p_{\perp} + \frac{B^{2}}{8\pi}\right)+ \nabla \cdot \left[\hat{\bm{b}} \hat{\bm{b}}\left( \Delta p + \frac{B^{2}}{4\pi}\right)\right],\label{eq:KMHD u}\\[2ex]
&\partial_{t} \bm{B} = \nabla \times (\bm{u} \times \bm{B} ),\label{eq:KMHD B} \\[2ex]
\partial_{t}p_{\perp} + \nabla \cdot (p_{\perp}& \bm{u}) + p_{\perp} \nabla \cdot \bm{u} + \nabla \cdot (q_{\perp}\hat{\bm{b}}) +q_{\perp} \nabla \cdot \hat{\bm{b}}\nonumber \\  & \qquad= p_{\perp} \hat{\bm{b}}\cdot (\hat{\bm{b}}\cdot \! \nabla \bm{u} )  -  \nu_{c}\Delta p,\label{eq:KMHD pp}
 \\[2ex]
\partial_{t}p_{\parallel} + \nabla \cdot (p_{\parallel}& \bm{u}) + \nabla \cdot (q_{\parallel}\hat{\bm{b}}) -2q_{\perp} \nabla \cdot \hat{\bm{b}} \nonumber \\  &\qquad = -2 p_{\parallel} \hat{\bm{b}}\cdot (\hat{\bm{b}}\cdot \! \nabla \bm{u} ) +2\nu_{c}\Delta p,\label{eq:KMHD pl}
\end{align}
where Gauss units are used, $\bm{u}$ and $\bm{B}$ are the plasma flow velocity and magnetic field,  $\rho$ is the mass density, $B\equiv \left| \bm{B} \right|$ and $\hat{\bm{b}}=\bm{B}/B$  denote the field strength and direction, and $q_{\perp}$ and $q_{\parallel}$ are heat fluxes along $\hat{\bm{b}}$ associated with the perpendicular and parallel thermal energies respectively.  We also define $\Delta \equiv \Delta p/p_{0}$ with $p_{0} = 2p_{\perp}/3+ p_{\parallel}/3$ (note $\Delta p \ll p_{0}$ for $\beta \gg 1$), and $v_{A} = B_{0}/\!\sqrt{4\pi\rho}$. 
While Eqs.~\eqref{eq:KMHD rho}-\eqref{eq:KMHD pl} will be solved numerically below (Fig.~\ref{fig:1}), in this section 
we make various approximations to 
 derive analytically the amplitude limits
and simplified wave equations. We consider two approximations for $\Delta p$ -- one collisionless ($\nu_{c}=0$), the other weakly collisional (Braginskii; $\Omega_{c}\gg \nu_{c}\gg |\nabla \bm{u}|$) --  neglecting
 compressibility in both cases (valid for $\beta \gg 1,\, \delta B_{\perp}/B_{0}\ll1$). 

When  $dB/dt<0$, the terms $ \hat{\bm{b}}\cdot (\hat{\bm{b}}\cdot \! \nabla \bm{u} )\approx B^{-1}dB/dt$ in Eqs. \eqref{eq:KMHD pp}-\eqref{eq:KMHD pl} locally force $\Delta = \Delta p /p_{0}<0$. Let us first consider collisionless ($\nu_{c}=0$) evolution of $\Delta $,
which is strongly influenced by heat fluxes for $\beta\gtrsim 1$. As a simple prescription for $q_{\perp,\parallel}$, we use a 
successful \emph{Landau fluid} (LF)  closure  \citep{Snyder:1997fs}, which (with $\Delta \ll 1$) posits
\begin{equation}
q_{\parallel} \approx -2\rho \sqrt{\frac{2 }{\pi }\frac{p_{\parallel} }{\rho }}\frac{k_{\parallel}}{| k_{\parallel}|} \left(\frac{p_{\parallel}}{\rho}\right), \: q_{\perp} \approx -\rho \sqrt{\frac{2 }{\pi }\frac{p_{\parallel} }{\rho }}\frac{k_{\parallel}}{| k_{\parallel}|} \left(\frac{p_{\perp}}{\rho}\right). \label{eq:heat fluxes}
\end{equation}
Further assuming $\hat{\bm{b}}\cdot \nabla q_{\perp,\parallel} \gg q_{\perp,\parallel} \nabla \cdot \hat{\bm{b}} $ (valid 
at $\delta B_{\perp}/B_{0} \ll 1$) and 
using $p_{\parallel}/\rho \approx p_{0}/\rho = c_{s}^{2}$, one obtains $\nabla \cdot (q_{\perp}\hat{\bm{b}}) +q_{\perp} \nabla \cdot \hat{\bm{b}} \sim -\rho c_{s}  |k_{\parallel}| (p_{\perp}/\rho)$ in the $p_{\perp}$ equation \eqref{eq:KMHD pp} [similarly for $p_{\parallel}$, Eq.~\eqref{eq:KMHD pl}]. This term, which models Landau damping of temperature perturbations, suppresses spatial variation in $p_{\perp,\parallel}$ over the particle crossing time  $\tau_{\mathrm{damp}}\sim (| k_{\parallel}| \,c_{s})^{-1}$. Thus, if $\tau_{\mathrm{damp}}\ll |\nabla \bm{u}|^{-1}$, the $k_{\parallel}\neq 0$ part of $\Delta$ is suppressed 
by $\sim v_{A}/c_{s}\sim \!\beta^{\,-1/2}$ compared to its mean, and a simple model is that $q_{\perp,\parallel}$
act to spatially average the $\Delta p $ driving, or 
\begin{equation}
\Delta = 3\!\int \!\left< \hat{\bm{b}}\cdot (\hat{\bm{b}}\cdot \! \nabla \bm{u} )\right>dt \left[1  + \mathcal{O}(\beta^{\,-1/2})(\bm{x}) \right]\approx  3 \left< \ln \frac{B(t)}{B(0)} \right>.\label{eq:noC Delta}
\end{equation}
Now consider the Braginskii limit, where collisions dominate ($\nu_{c}\gg \nabla \bm{u}$). 
Equations~\eqref{eq:KMHD pp} and \eqref{eq:KMHD pl} then give
\begin{equation}
\Delta \approx {\nu_{c}}^{-1} \hat{\bm{b}}\cdot (\hat{\bm{b}}\cdot \! \nabla \bm{u} ),\label{eq:brag Delta}\end{equation}
neglecting
$q_{\perp,\parallel}$ for simplicity (valid for $\delta p_{\perp,\parallel} /p_{\perp,\parallel} \ll |\bm{u}|/c_{s}$).

Furnished with approximations for $\Delta p$  [Eq.~\eqref{eq:noC Delta} or \eqref{eq:brag Delta}], we now examine SA fluctuation dynamics.
Consider a 
background field $B_{0}\hat{\bm{z}}$, with perturbations perpendicular to $\hat{\bm{z}}$ and the wavevector $\bm{k} = k_{\parallel} \hat{\bm{z}}+ \bm{k}_{\perp}$. Since SA waves are  unmodified by $\bm{k}_{\perp} \neq 0$  (the envelope is simply modulated in the perpendicular direction) and we analyze linear polarizations,  we  take $x$-directed
perturbations that depend only on $z$ and $t$;
$\bm{B}=B_{0}\,\hat{\bm{z}} + \delta {B}_{\perp}(z,t)\, \hat{\bm{x}}$,  $\bm{u} = {u}_{\perp}(z,t)\, \hat{\bm{x}}$.
Neglecting compressibility, the  field perturbation $\delta b = \delta B_{\perp}/B_{0}$ satisfies
\begin{equation}
\frac{\partial^{2}}{\partial t^{2}}\delta b = v_{A}^{2}\left[ \frac{\partial^{2}}{\partial z^{2}} \delta b + \frac{\beta}{2} \frac{\partial^{2}}{\partial z^{2}} \left(\frac{\delta b}{1+\delta b^{2}} \Delta(z) \right)\right]. \label{eq:waves}
\end{equation}
Equation~\eqref{eq:waves} illustrates that in the absence of a background $\Delta$ linear long-wavelength
SA fluctuations are unmodified by kinetic effects, while the parallel firehose 
occurs because the coefficient of $\partial_{z}^{2}(\delta b) $ is 
negative for $\beta \Delta/2 < -1$.

Combining Eqs.~\eqref{eq:noC Delta} and \eqref{eq:waves}, we see that if a collisionless wave evolves such that  $3 \left<\ln [B(t)/B(0)] \right>  = -2/\beta $,  its restoring force disappears. 
As we now explain, although the amplitude limit in each case is the same, standing and traveling waves differ in \emph{why} a decrease in $\langle B(t) \rangle$ occurs.
In a standing wave starting from a magnetic perturbation, $\langle B(t) \rangle$ simply decreases as the wave evolves. Thus if $-3 \left<\ln [B_{0}/B(0)] \right> \approx 3 \delta b(0)^{2}/4 > 2/\beta $ [assuming $\delta b(0) \!\sim \!\delta b_{0}\sin (k_{\parallel }z) \ll 1$], 
an interruption occurs before a quarter period, implying the maximum wave amplitude is
\begin{equation}
\left( \frac{\delta B_{\perp}}{B_{0}}\right)_{\mathrm{max}} \approx \sqrt{\frac{8}{3}}\, \beta^{\,-1/2}.\label{eq:noC condition}\end{equation}
This limit is matched nearly perfectly by numerical LF solutions (see Fig.~\ref{fig:2}).
 A standing wave with an initial velocity perturbation also satisfies the limit~\eqref{eq:noC condition} and is addressed in more detail below. 
For traveling waves, a crucial role is played by the spatially dependent $\mathcal{O}(\beta^{\,-1/2})$ part of $\Delta$,
which we neglected for convenience in deriving Eq.~\eqref{eq:noC condition}. This role is to decrease $\langle B(t) \rangle$ 
by damping the wave nonlinearly. This ``pressure-anisotropy damping'' is related to  correlations between 
$B^{-1}dB/dt$ and $\Delta p$, which cause a contribution to the rate of change of thermal
energy of the 
 form $\sim\!\!\int d\bm{x} \,\Delta p\, B^{-1}dB/dt $ [see Eqs.~\eqref{eq:KMHD B}--\eqref{eq:KMHD pl}]. 
Because this is positive for a traveling wave,  the wave heats the plasma and damps at the rate $\sim\! \omega_{A}\delta b^{2} \beta^{\,1/2} $ (where $\omega_{A}=k_{\parallel}v_{A})$. Without any mechanism to isotropize the pressure, the decrease in $\langle B(t) \rangle$ causes  $\langle \Delta\rangle $ to decrease as $\langle\Delta\rangle= 3 \left<\ln [B(t)/B(0)] \right>$ [Eq.~\eqref{eq:noC Delta}], which slows the wave [see Fig.~\ref{fig:1}(c)] before stopping it completely if $\langle \Delta\rangle = -2/\beta$.
The maximum amplitude of a traveling wave is thus also given by Eq.~\eqref{eq:noC condition}, although the time to approach the limit is increased  compared to the standing wave because of the time required
for the wave to damp nonlinearly. 

A similar estimate of the amplitude limit with the Braginskii closure \eqref{eq:brag Delta}, using
$\partial_{t}(\delta b) \sim \omega_{A} \delta b$, yields
\begin{equation}
\frac{\beta}{2}\frac{\omega_{A}}{\nu_{c}} \delta b(0)^{3} \lesssim \delta b(0)\quad \Longrightarrow \quad \delta b(0)_{\mathrm{max}} \sim \sqrt{\frac{\nu_{c}}{\omega_{A}}} \beta^{\,-1/2}.\label{eq:brag condition}\end{equation}
Since ${\nu_{c}}/{\omega_{A}}\gg1$ for the validity of Braginskii's approximation, this condition is less stringent 
than Eq.~\eqref{eq:noC condition}; note also that it depends on $k_{\parallel}$ (via $\omega_{A}$) unlike the collisionless 
case. 
In the Braginskii limit, traveling waves are again strongly damped [at the rate $\sim \! (\omega_{A}/\nu_{c})\, \delta b^{2} \beta\,\omega_{A}$] due to  spatial correlation of
 $\Delta$ and $B^{-1}dB/dt$.


\section{Nonlinear evolution and numerical results}
The results above naturally invite the question: what happens to fluctuations above the critical amplitude? Here we illustrate,
through numerical solutions and simple arguments, the  
remarkable tendency of collisionless plasmas to minimize the variation in $B^{2}$ \citep{Kunz:2014kt,Rincon:2015tq,Melville:2015tt,Melville:2016tbc}. As a result, an initially sinusoidal $\delta b$ relaxes 
 into a square wave,  corresponding to  zig-zags in the
 field lines. This peculiar behavior also emerges
from Eqs.~\eqref{eq:noC Delta} and \eqref{eq:waves}, despite their simplicity, illustrating the effect's simple physical origins. Solutions using Braginskii 
MHD differ in appearance and damp to fluctuations with $\delta b < (\delta b)_{\mathrm{max}}$.

We solve equations \eqref{eq:KMHD rho}-\eqref{eq:KMHD pl} with the LF closure, using a dealiased pseudo-spectral method and hyperviscously damping all variables to remove energy just above the grid scale. Our only further approximation is the identification of $1/| k_{\parallel} |$ in Eq.~\eqref{eq:heat fluxes} with $1/| k_{z}| $ (valid for $\delta b \ll 1$).
The full equations solved are Eqs.~(35)--(44) of \citet{Sharma:2006wp} [except we use $1/| k_{z}| $ in Eq.~\eqref{eq:heat fluxes}, not their
$k_{L}$]. We do not artificially limit the pressure anisotropy to the firehose limit, as common in previous
turbulence studies  \citep{Sharma:2006wp,SantosLima:2014cn}. This is because the parallel firehose instability -- the cause of the effect -- is in fact captured by the fluid model. In addition, since finite Larmor radius effects (FLR) are not contained in this LF model, all scales in the simulation are  larger than the 
gyroradius.\footnote{Firehose fluctuations are damped due to hyperviscosity, which thus determines the scale
of the fastest growing firehose modes.}

The evolution of a sinusoidal SA perturbation is shown
in Fig.~\ref{fig:1}, starting with a perturbation in either (a) $\bm{B}$, (b) $\bm{u}$, or (c) a traveling wave. For comparison, we show solutions of the nonlinear wave equation \eqref{eq:waves} in panels (d)--(e).
In (b)--(c), $\Delta$ is limited at the mirror threshold $\Delta=1/\beta$, since  $dB/dt>0$ in some regions (see discussion below). We see from Fig.~\ref{fig:1}(a)--(d) that collisionless waves -- both standing and traveling -- generically relax  to a stable sequence of near-perfect stair steps. The spatial scale of the jumps
 is set by the numerics, so would likely be determined by FLR effects in reality.
The basic origin of such structures can be understood by observing that if $\langle \Delta \rangle= -2/\beta $ in Eq.~\eqref{eq:waves}, 
$\partial^{2}_{z}(\delta b)+(\langle\Delta\rangle \beta/2)\, \partial^{2}_{z} (\delta b)=0$. Neglecting residual spatial variation in $\Delta$ (this decreases after wave interruption because $B^{-1}dB/dt$ decreases), the remainder on the right-hand side of Eq.~\eqref{eq:waves} is $\sim \!\partial_{z}^{2}(\delta b^{3})$, which lowers  maxima of $\delta b^{2}$ while increasing minima, leading to constant-$B$ steps.  
With the Braginskii closure [Fig.~\ref{fig:1}(e)], in contrast to the collisionless case, regions of small $\delta b$ have smaller $|\Delta |$ and thus decrease to zero before $\Delta=-2/\beta$.
Further, since the nonlinearity is diffusive, the field decays (over the timescale $ \tau_{\mathrm{decay}} \sim \beta\, \delta b(0)^{2}/\nu_{c}$),\footnote{ This estimate for $\tau_{\mathrm{decay}}$ can be derived by setting $\Delta$ [Eq.~\eqref{eq:brag Delta}] equal to the firehose limit in Eq.~\eqref{eq:waves} and solving the resulting differential equation. It is well matched by numerical solutions. } leaving
 small $\delta b< (\delta b)_{\mathrm{max}}$ fluctuations.

 \begin{figure}
\begin{center}
\includegraphics[width=1.0\columnwidth]{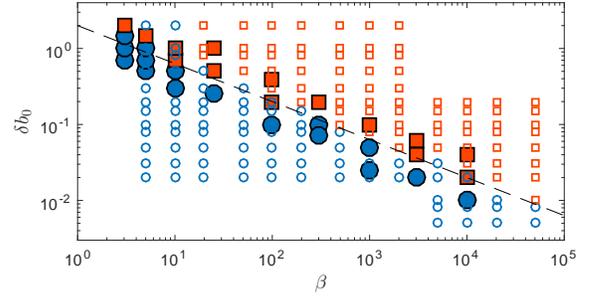}
\caption{Numerical confirmation of the scaling \eqref{eq:noC condition}. A red square indicates that
 an initial magnetic perturbation was interrupted before a half cycle (as in Fig.~\ref{fig:1}), while a blue circle indicates that the perturbation flipped polarity without interruption. Large filled symbols show results from the LF equations \eqref{eq:KMHD rho}-\eqref{eq:KMHD pl}, while small hollow symbols 
show solutions of Eq.~\eqref{eq:waves} with the collisionless closure \eqref{eq:noC Delta}. The dashed 
line is $\delta b_{0} = 2 \beta^{\,-1/2}$.
}
\label{fig:2}
\end{center}
\end{figure}
Figure \ref{fig:2} confirms the predictions of Eq.~\eqref{eq:noC condition}, illustrating  essentially perfect agreement for $\beta \gtrsim 10$.
 At $\beta \lesssim 10$  large-amplitude waves are still interrupted in the LF model, although solutions of
Eq.~\eqref{eq:waves} (which required $\delta b\ll 1$) deviate from Eq.~\eqref{eq:noC condition}. We have also confirmed the scaling \eqref{eq:brag condition} for Braginskii 
MHD (not shown).

So far we have  considered only 1-D evolution within the LF model -- what caveats should be applied for more realistic conditions?  The reader may wonder about the imposition of a mirror (but no firehose) 
limit in Fig.~\ref{fig:1}(b)--(c). This is required because our model cannot 
capture the mirror instability, which gives rise to growing modes at $k_{\perp}\gg k_{\parallel}$. 
However, kinetic results \citep{Kunz:2014kt,Rincon:2015mi,Hellinger:2015mi,Melville:2015tt}  
show that
 mirror fluctuations limit $\Delta$ by trapping particles, allowing $B$ to continue increasing while maintaining $\Delta = 1/\beta$. Further, the temporal growth  of the mirror instability 
 $|\delta B_{\parallel}/B_{0}|\sim (|\nabla \bm{u}|t)^{2/3 }$ \citep{Rincon:2015mi} is slow enough that mirrors generated by a SA wave will not saturate and significantly scatter particles if $u_{x}(0)/v_{A}<1$.
 Thus, following a $\bm{u}$ perturbation with $B^{-1}dB/dt>0$, mirrors grow to  limit $\Delta=1/\beta$; then, once $B^{-1}dB/dt<0$, $\Delta$ immediately starts decreasing, while the (small) mirror fluctuations decay at the rate $\gamma \sim \Omega_{c}/\beta$ \citep{Melville:2015tt}. This implies that SA waves cannot circumvent the limit \eqref{eq:noC condition} by starting from $B=0$ or $\Delta>0$ [see Fig.~\ref{fig:1}(b)].
Oblique firehose fluctuations \citep{Yoon:1993of,Hellinger:2008ob} are also not captured by our model, and these may 
change the nonlinear behavior by scattering particles \citep{Kunz:2014kt}, potentially disrupting the angular field structure.\footnote{The angular magnetic structures themselves may 
also scatter particles, with $\nu_{c}\sim k_{\parallel}c_{s}$. This could 
cause faster damping of a wave once it hits interruption limit and becomes square [or perhaps earlier for traveling waves, which can become square before $\Delta=-2/\beta$; see Fig.~\ref{fig:1}(c)].} Again, however, they cannot circumvent the amplitude
limit itself; they become active only once $\Delta < -2/\beta$, when the wave restoring force has already disappeared.
We thus stress that, although the nonlinear 
outcome of wave interruptions (Fig.~\ref{fig:1}) may be modified by the addition of other kinetic physics, 
our basic result -- that collisionless SA fluctuations cannot exist in their linear wave form above the limit \eqref{eq:noC condition} -- 
is robust. Its derivation is insensitive to details of heat fluxes or particle scattering at the microinstability boundaries, relying purely on the physics of pressure anisotropy generation due to magnetic
moment conservation.

\section{Implications}
Given the ubiquity of Alfv\'en waves in space and astrophysical  plasmas, the implications of the stringent 
constraint  \eqref{eq:lim} on their amplitude at high $\beta$ may be dramatic, with  applications ranging from the intracluster medium \citep{Zhuravleva:2014}, to hot (collisionless) accretion disks \citep{Quataert:1999gn} and 
the solar wind near Earth \citep{Bruno:2013hk}. We leave much of the discussion of these applications to future work, briefly considering possible observational evidence for the effect in the  solar wind and the implications for magnetized turbulence \citep{Goldreich:1995hq}.
Note that, in contrast to results presented here,  linear damping of long-wavelength, low-frequency SA waves at high $\beta$ is negligible if $\Omega_{c}\, \beta^{-1}/\omega_{A} \gg1$ \citep{Foote:1979ee,Achterberg:1981,Cramer:2011uc}.

Alfv\'en waves are fundamental to solar wind physics, and our results 
are most relevant to regions where $\beta\gtrsim 1$,   at solar radii $\sim\!\!1 \mathrm{AU}$ \citep{Mullan:2006,Bruno:2013hk}. Specifically,  propagation of large-amplitude SA waves into  a  $\beta\gtrsim 1$ plasma
may naturally form rotational field-line discontinuities  \citep{Borovsky:2008eo,Miao:2011hv}, heating the plasma as the wave interrupts. 
An interesting observational feature that may be related to this
is the appearance of a distinct, magnetically dominated, population of fluctuations at increasing solar radii \citep{Tu:1991,Bruno:2007,Bruno:2013hk}. 
This population's sudden appearance across a range of latitudes \citep{Bavassano:1998} suggests it does not arise through continuous
evolution of turbulence (see Fig.~2 of \citealt{Bruno:2007}). Such characteristics would be expected from
SA wave interruption in regions where $\beta\gtrsim 1$, a scenario that is also consistent with the observed excess of magnetic energy \citep{Goldstein:1995,Roberts:2010bf,Chen:2012sw,Oughton:2015cv}. 
A prediction of our scenario is a correlation
between $\beta$ and regions with magnetically dominated, rotationally discontinuous, structures.

The implications of our results for magnetized  turbulence in collisionless plasmas are potentially dramatic. 
A striking conclusion, which holds independently of the details of interrupted structures, is that perturbations in a collisionless plasma 
with energy densities on the order of  $B_{0}^{2}$ (i.e., $|\bm{u}|\sim v_{A}$)  are immediately damped  -- that is, the plasma behaves as a fluid with Reynolds number $\lesssim \!1$. 
Where does this perturbation energy go? 
Because of the same energy transfer term responsible for damping traveling waves, $\partial_{t}E_{\mathrm{th}}\sim \int d\bm{x} \,\Delta p\, B^{-1}dB/dt$,
if $\Delta<0$, a decreasing field   
\emph{directly} transfers large-scale kinetic energy into plasma heating \citep{Sharma:2006wp}. 
A turbulent cascade is thus no longer necessary for collisionless
plasmas to absorb the energy  input by a continuous mechanical forcing \citep{Kunz:2010gv}, and
it is unclear if any of the energy provided on large scales cascades to smaller scales as traditionally assumed. 
However, such physics is well beyond the scope of this work and we  conclude here by simply reiterating
that the immediate disruption of SA fluctuations when $\delta {B}_{\perp}/{B}_{0} \gtrsim \beta^{\,-1/2}$ 
severely limits the application of standard magnetized turbulence phenomenologies \citep{Goldreich:1995hq} to  high-$\beta$ collisionless plasmas.

A variety of fundamental questions about the nonlinear interruption of shear-Alfv\'en waves remain for future studies,
particularly concerning higher-dimensional microinstabilities (e.g., oblique firehose).
Fully kinetic simulations will be key to understanding this physics better.
Given the robustness and generality of our result, its appearance in a variety of models, and the stringent nature 
of the $\delta {B}_{\perp}/{B}_{0} \lesssim \beta^{\,-1/2}$ condition, we anticipate a
 range of future applications to heliospheric, astrophysical, and possibly laboratory \citep{Forest:2016,LAPD:2016} plasmas.

\begin{acknowledgments}
It is a pleasure to thank S.~Bale, C.~H.~K~Chen, S.~Cowley, M.~Kunz, and M.~Strumik for useful discussions. This work was supported in part by the Sherman Fairchild foundation (JS), the Gordon and Betty Moore Foundation to Lars Bildsten, Eliot Quataert and E. Sterl
Phinney, Simons Investigator awards from the Simons Foundation (EQ), NSF grant AST 13-33612 (EQ), and by UK STFC and EPSRC grants (AAS). 
\end{acknowledgments}


\end{document}